\documentclass[aps,pre,twocolumn,showpacs,superscriptaddress,10p,longbibliography]{revtex4-1}

\usepackage{amsmath}
\usepackage{amssymb}
\usepackage[colorlinks=true,linkcolor=red]{hyperref}
\usepackage[sort&compress]{natbib}
\usepackage{graphicx}
\usepackage{color}

\begin{document}

\title{Electromagnetic waves propagation through an array of superconducting qubits: manifestations of non-equilibrium steady states of qubits}

\author{M. V. Fistul}
\affiliation{Center for Theoretical Physics of Complex Systems, Institute for Basic Science (IBS), Daejeon 34051, Republic of Korea} \affiliation{National University of Science and Technology "MISIS", Russian Quantum Center  119049 Moscow, Russia}

\author{M.A.  Iontsev}
\affiliation{National University of Science and Technology "MISIS", Russian Quantum Center 119049 Moscow, Russia}

%\author{S. I. Mukhin}
%\affiliation{National University of Science and Technology %"MISIS", Russian Quantum Center 119049 Moscow, Russia}

\date{\today}

\begin{abstract}
 We report a theoretical study of the electromagnetic waves (EWs) propagation through an array   of  superconducting qubits, i.e. coherent two-level systems, embedded in a low-dissipative transmission line.  We focus on the near-resonant case as the frequency of EWs $\omega \simeq \omega_q$, where $\omega_q$ is the qubit frequency. In this limit we derive the effective dynamic nonlinear wave equation allowing one to obtain the frequency dependent transmission coefficient of EWs, $D(\omega)$. In the linear regime and  a relatively wide frequency region we obtain a strong resonant suppression of $D(\omega)$ in  both cases of a single qubit and  chains composed of a large number of densely arranged
 qubits. However, in narrow frequency regions a chain of qubits allows the resonant transmission of EWs with greatly enhanced $D(\omega)$. In the nonlinear regime realized for a moderate power of applied microwave radiation, we predict and analyze various transitions between states characterized by high and low values of $D(\omega)$. These transitions are manifestations of nonequilibrium steady states of an array of qubits  achieved in this regime. 
\end{abstract}
%\pacs{42.50.-p,74.81.Fa,74.50.+r}

\maketitle

\section{Introduction}
Electromagnetic waves (EWs) propagation in metamaterials-artificially prepared media composed of a network of interacting lumped electromagnetic circuits- attracts recently an enormous attention  due to variety  of physical phenomena occurring  in such systems, e.g. electromagnetically induced transparency (reflectivity) \cite{liao2016electromagnetically,shulga2018magnetically,chaldyshev2011optical}, "left-handed" metamaterials \cite{smith2004metamaterials,zharov2003nonlinear}, dynamically induced
 metastable states \cite{lazarides2015chimeras,jung2014multistability}, just to name a few. These networks have been fabricated from metallic, semiconducting, magnetic or superconducting materials. 

The latter case of networks based on superconducting elementary circuits  presents a special interest because of an extremely low dissipation, a great tunabilty of the microwave resonances, and a strong nonlinearity \cite{jung2014multistability,ricci2005superconducting,anlage2010physics}. In most of studied systems these superconducting electromagnetic circuits contained by one or a few Josephson junctions, can be precisely described as  classical nonlinear oscillators, and the interaction of propagating EWs with a network of such superconducting  lumped circuits is determined by a set of classical nonlinear dynamic equations \cite{jung2014multistability,miroshnichenko2001breathers,filatrella2000high}. 

However, it is  well known for many years that small superconducting circuits can be properly biased in the \textit{coherent macroscopic quantum regime}, and in a simplest case the dynamics of such circuits is equivalent to the quantum dynamics of two-level systems, i.e. qubits \cite{pashkin2003quantum,houck2009life,chiorescu2004coherent,majer2007coupling,fink2009dressed,macha2014implementation,shulga2017observation}. A surfeit of different types of superconducting qubits  has been realized , e.g. dc voltage biased charge qubits (Fig. 1A) \cite{pashkin2003quantum}, flux qubits weakly \cite{macha2014implementation} (Fig. 1B) and strongly (Fig. 1C) \cite{shulga2018magnetically} interacting with a low-dissipative transmission line, transmons \cite{houck2009life,shulga2017observation} etc. 

As a next step these qubits are organized in different arrays or lattices forming \textit{ quantum electromagnetic networks}, and an inductive or capacitive coupling of such networks to an external low-dissipative transmission line allows one to experimentally access the frequency dependent transmission coefficient of EWs, $D(\omega)$.  The interaction of EWs with quantum networks of qubits results in a large amount of coherent quantum phenomena on a macroscopic scale , e.g. collective quantum  states \cite{fink2009dressed,macha2014implementation,shulga2017observation,fistul2017quantum}, magnetically induced transparency \cite{shulga2018magnetically} have been observed, and the coherent electromagnetic pulses propagation \cite{asai2015effects,ivic2016qubit}, the nonclassical states of photons \cite{iontsev2016double} have been theoretically predicted and studied. 
Therefore, in this quickly developing field a natural question arises \cite{rakhmanov2008quantum,asai2015effects,ivic2016qubit,fistul2017quantum,iontsev2016double}: how the coherent quantum dynamics of a network of superconducting qubits influences the EWs propagation? 

In this paper we present a systematic study of EWs propagation through an array of qubits embedded in a low-dissipative transmission line (see Fig. 1). We will focus on the resonant case, i.e. $\omega \simeq \omega_q$, where $\omega_q$ is the qubit frequency, and the transmission coefficient $D(\omega)$ will be theoretically analyzed. To obtain $D(\omega)$ we derive the effective nonlinear EWs equation taking into account the coherent quantum dynamics of qubits exposed to the electromagnetic field. It allows us to address both limits, low and high power of applied microwave radiation. 

In the linear regime and  a relatively wide frequency region near the resonance we obtain a strong suppression of $D(\omega)$ in  both cases of a single qubit and chains composed of a large number of densely arranged qubits. However,  in a narrow frequency region for chains of qubits  we obtain the resonant transmission of EWs with a greatly enhanced $D(\omega)$. As we turn to the nonlinear regime  realized for a moderate power of applied microwave radiation, we predict and analyze various transitions between states characterized by high and low values of $D(\omega)$. We argue that these transitions are fingerprints of nonequilibrium steady states of an array of qubits.

%    Figure 1
\begin{figure}
    \includegraphics[width=1\columnwidth]{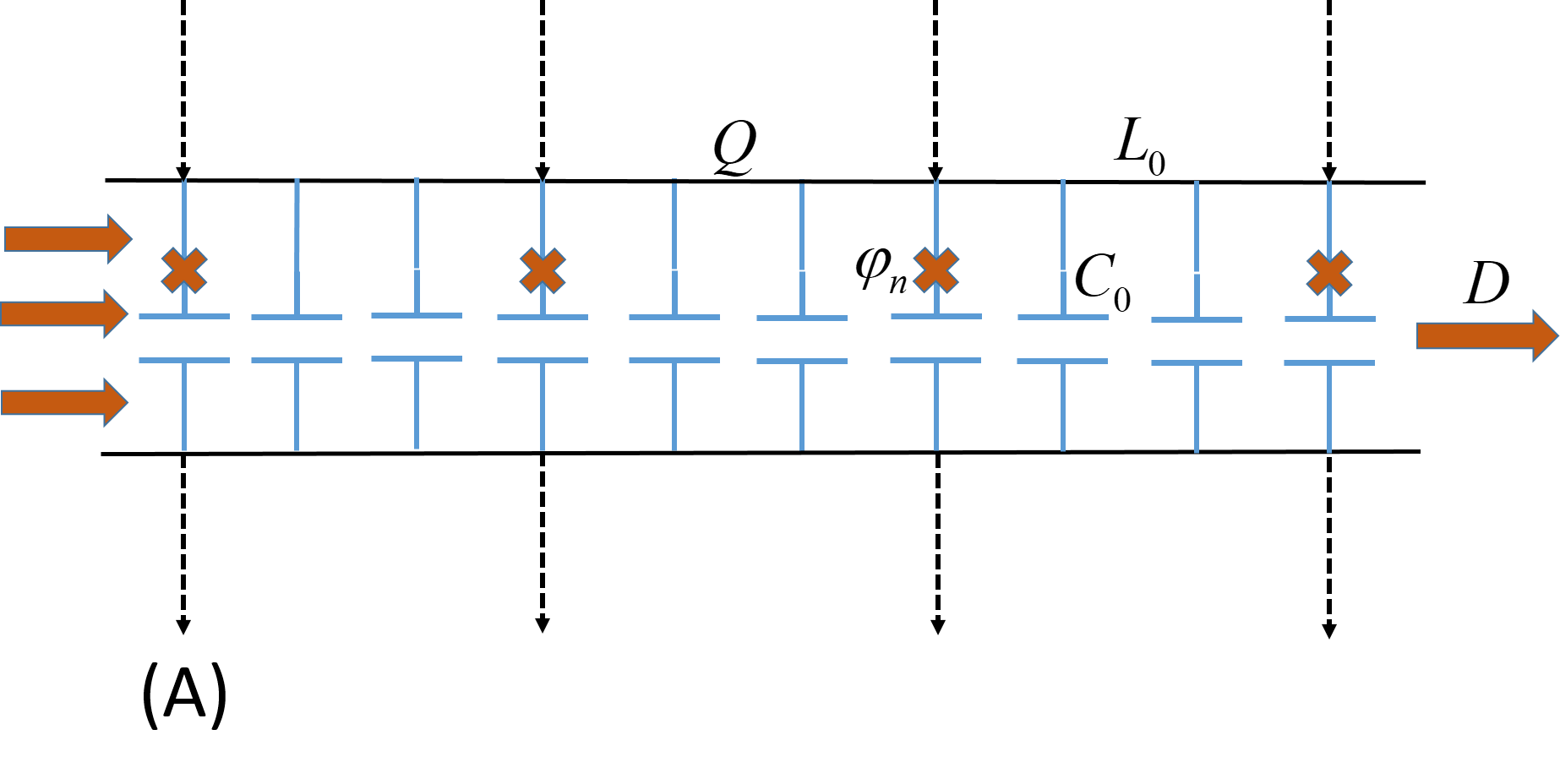}
    \includegraphics[width=1\columnwidth]{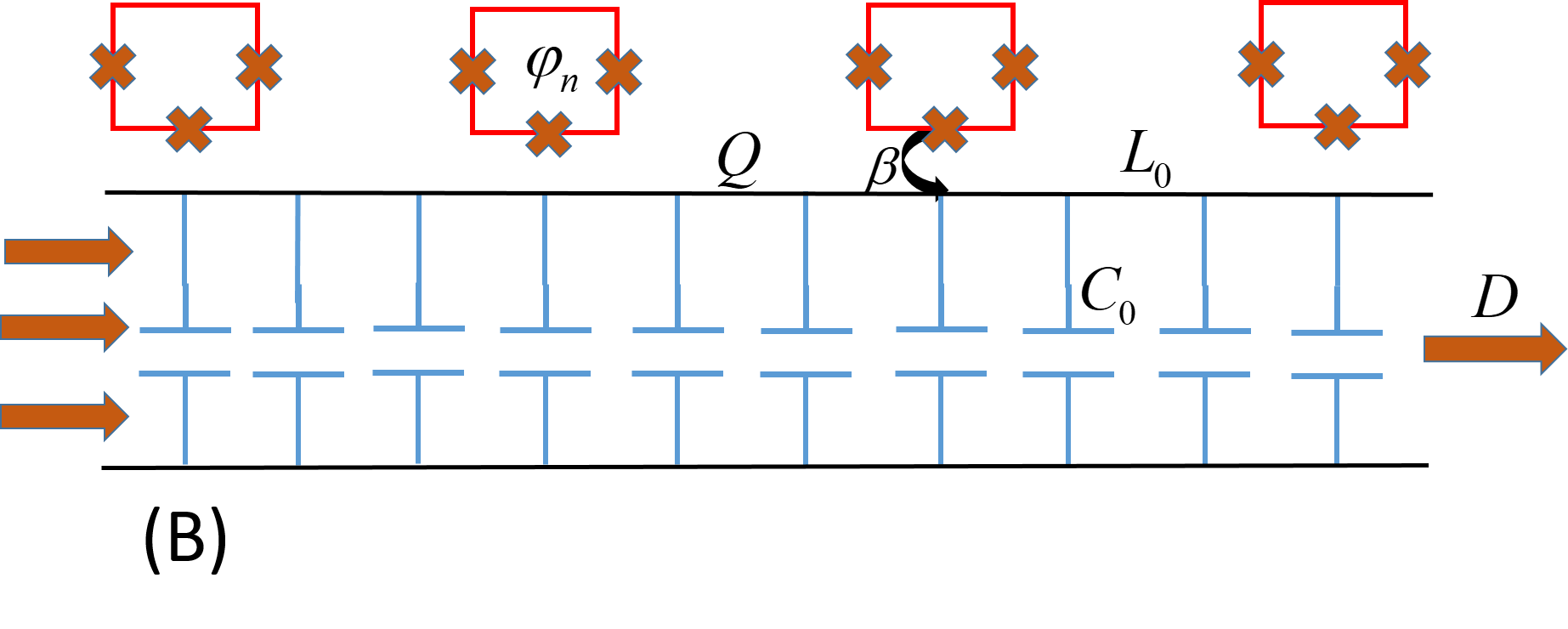}
    \includegraphics[width=1\columnwidth]{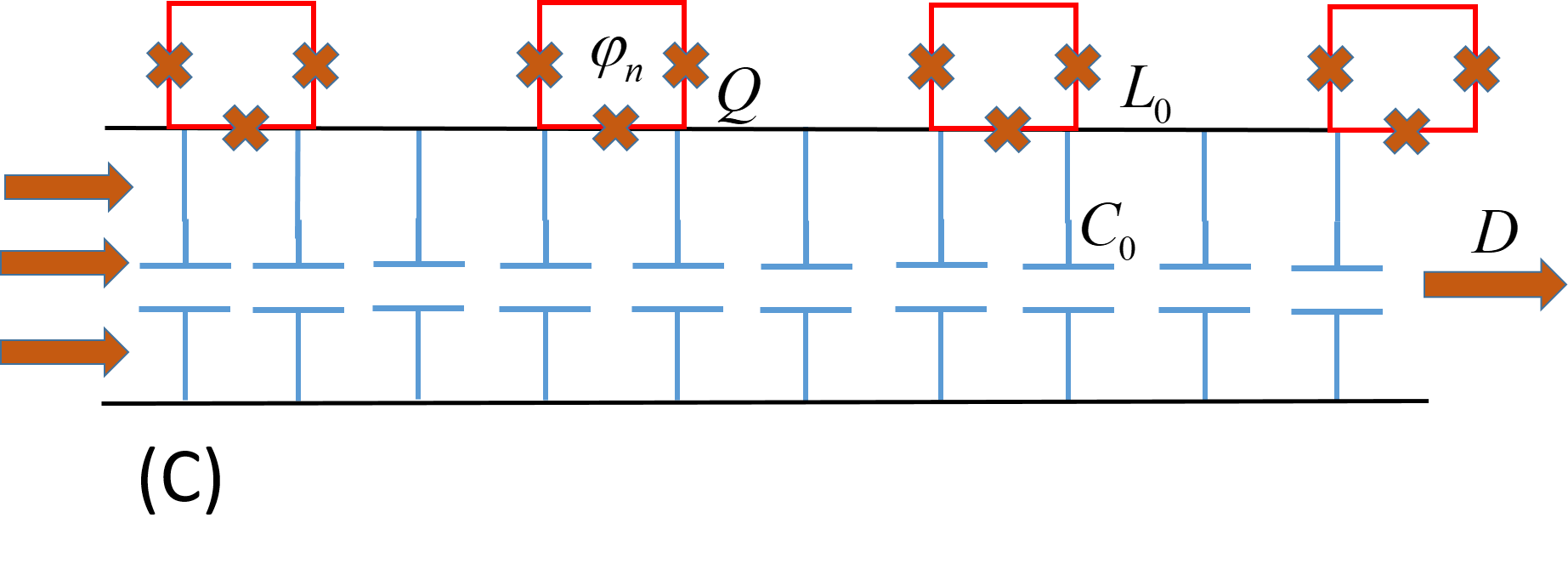}
    \caption{(color online) Schematic of qubit arrays coupled to a low-dissipative transmission line: voltage biased charge qubits (A), weakly coupled flux qubits (B) and strongly coupled flux qubits (C). Josephson junctions are illustrated by crosses, imput and output of EWs are shown by arrows. The classical, $Q$, and quantum $\varphi_n$ dynamic variables are shown. The properties of transmission line are characterized by two parameters:  the capacitance $C_0$ and inductance $L_0$  per length. The $D$ is the transmission coefficient of propagating EWs. }
    \label{fig:schematic}
\end{figure}

The paper is organized as follows: In Section II we present our model for qubits array embedded in a low-dissipative transmission line, introduce the Lagrangian,  and derive the effective  nonlinear wave equation for the electromagnetic field interacting with an array of qubits. In Sec. III we analyze the coherent quantum dynamics of a single qubit subject to an applied electromagnetic field in both limits of low and high power. In Sec. IV we apply derived in Sec. II the effective nonlinear wave equation to a study of frequency dependent transmission coefficient, $D(\omega)$, for  a chain of densely arranged qubits. Moreover, we address both regimes, i.e. linear and nonlinear ones.   The Section V provides conclusions. 

\section{Model, Lagrangian and Dynamic Equations}

\subsection{Model}

Let us consider a regular one-dimensional  array of $N$ lumped superconducting quantum circuits embedded in a low-dissipative nondispersive transmission line (see Fig. 1). As the amplitude of propagating EWs is not too low, i.e in the regime of large number of photons, the electromagnetic field in the transmission line is characterized by  coordinate and time dependent classical variables - the charge distribution, $Q(x,t)$. Different types of lumped superconducting quantum circuits have been realized (the schematics of arrays composed of charge (Fig. 1(A)) and flux (Fig. 1(B),(C)) qubits are shown),  and the quantum dynamics of such circuits is characterized by  quantum variables-the Josephson phases, $\varphi_n$. An artificially prepared potential $U(\varphi_n)$ allows one to vary the circuits resonant frequencies in a wide region.   
The dynamics of a whole system in the classical regime is described by total Lagrangian, which consists of three parts: the Lagrangian of electromagnetic field $L_{EF}$,  the Lagrangian of an array of lumped superconducting quantum circuits (qubits) $L_{qb}$, and the interaction Lagrangian $L_{int}$ describing the interaction between qubits and electromagnetic field:
\begin{equation} \label{Lagrangian}
L=L_{EW}+L_{qb}+L_{int}.
\end{equation}

\subsection{Lagrangian and dynamic equation}
The electromagnetic field Lagrangian $L_{EF}$ is written as
\begin{equation}
    \label{Lagph}
    \ L_{EF}=\frac{L_0 \ell}{2} \left \{ \left [\frac{\partial Q}{\partial t} \right ]^2-c_0^2 \left [\frac{\partial Q}{\partial x} \right]^2 \right \}, 
\end{equation}
where $c_0=1/\sqrt{L_0C_0}$ is the velocity of EWs propagating in the transmission line, $L_0$ and $C_0$ are the inductance and capacitance of the transmission line per length, accordingly;  $\ell$ is the length of the system. 

The Lagrangian of an array of lumped quantum circuits is written as

\begin{equation}
    \label{Lagj}
    \ L_{qb}=  \sum_{n=1}^N \frac{E_J}{2\omega_p^2}(\dot{\varphi}_n-\dot{\phi}_{0n})^2-U(\varphi_n),
\end{equation}
where $E_J$, $\omega_p$ are the Josephson energy and the plasma frequency, accordingly; the parameter $\dot{\phi}_{0n}$ is proportional to the gate voltage, and it allows one to vary the frequency of charge qubits (such gate circuits are shown by dashed arrows in Fig. 1(A)). For charge qubits the potential $U(\varphi_n)$ is written explicitly as $U(\varphi_n)=E_J (1-\cos \varphi_n)$ whereas for flux qubits (see Fig. 1(B),(C)) the double well potential reads as: $U(\varphi_n)=-E_J [2\cos \varphi_n-\eta \cos (2\varphi_n)]$.

The interaction part of the Lagrangian is written as
\begin{equation}
    \label{Lagint}
    \ L_{int}= - \frac{\hbar\alpha w }{2e}\sum_{n=1}^N \delta (na-x) Q(t,x)\dot{\varphi}_n,
\end{equation}
where the coupling coefficient $\alpha$ varies by around five order of magnitude from $10^{-2}$ for a weak coupling between qubits and transmission line \cite{macha2014implementation} (Fig. 1(B)) up to $4 \times 10^2$  for  strongly coupled qubits \cite{shulga2018magnetically} (Fig. 1(C)). The $w$ is the geometrical size of lumped quantum circuits (qubits), $w \ll \ell$.  

As we turn to the coherent quantum regime of networks of qubits, the dynamics of EWs is described by the specific wave equation as
\begin{equation}
    \label{dynequation}
   \frac{\partial^2 Q}{\partial t^2} -\gamma \frac{\partial Q}{\partial t} -c_0^2 \frac{\partial^2 Q}{\partial x^2} = -\frac{  \hbar}{2e} \frac{ \alpha w}{ L_0 \ell}\sum_{n=1}^N \delta (na-x) <\dot{\varphi}_n>_{eq},
\end{equation}
where  we take into account the dissipation effects in the transmission line characterized by phenomenological parameter $\gamma \ll 1$. Here, $<...>_{eq}$ denotes the quantum mechanical averaging over the equilibrium state of quantum network.

\subsection{Quantum dynamics of a single qubit and effective wave equation}
Since we neglect the direct coupling between elementary circuits, the coherent quantum dynamics of a network is reduced to the sum of independent lumped electromagnetic circuit exposed to an applied electromagnetic field.
The quantum dynamics of a single element is determined by the time-dependent Hamiltonian, $\hat H_{qb}=\hat H_0+\hat H_{t}$, where
the equilibrium Hamiltonian $\hat H_0$ is
\begin{equation}
    \label{eqHamil}
 \hat H_0=   \frac{\omega_p^2}{2E_J}(\hat p_\varphi-p_0)^2+U(\varphi)
\end{equation}
and the nonequilibrium part of the total Hamiltonian, explicitly depending on time, $\hat H_t$ is
\begin{equation}
    \label{neqHamil}
 \hat H_t=  \frac{\hbar \alpha \omega_p^2}{2e E_J} Q(t,x) \hat p_\varphi.
\end{equation}
In the resonant regime as the EW frequency $\omega \simeq \omega_q$, we truncate the explicit Hamiltonian (see Eqs . (\ref{eqHamil}) and (\ref{neqHamil})) to the Hamiltonian of two-levels systems. These two levels can be fine-tuned to the resonance with the frequency of EW propagating in the transmission line.  In particular, for the charge qubits case shown in  Fig. 1(A), on the avoid-crossing point such truncation leads to the effective Hamiltonian written as \cite{pashkin2003quantum}
\begin{equation} \label{effHamiltonian}
   \hat H_{eff}=\frac{E_J}{2} \hat \sigma_z+\frac{(\hbar \omega_p)^2 \alpha}{4e E_J} Q(x,t)\hat \sigma_x,
\end{equation}
where $\hat \sigma_{x,z}$ are the Pauli matrices. In this case the qubit frequency is expressed as $\omega_q=E_J$. The time-dependent wave function of a charge qubit is written as
\begin{equation}
    \Psi (t)=C_-(t)f_-+C_+(t)f_+,
\end{equation}
where $f_{\pm}=\frac{1}{\sqrt{4\pi}}\left(1\pm e^{i\varphi} \right)$ stationary wave functions of two states. The corresponding quantum-mechanical average of the operator $<\dot \varphi>$  in the right hand part of Eq. (\ref{dynequation}) reads as
\begin{equation} \label{Average}
 <\dot \varphi>_{eq}= \frac{\hbar \omega_p^2}{E_J}\Re e [C_-(t)C_+(t)]
\end{equation}
Taking into account the initial conditions  $C_{-}(0)=1$ and $C_{+}(0)=0$, and using the resonant condition, $\omega_q \simeq \omega$  we obtain in the non-dissipative (\textit{nd}) regime
$$
 S^{nd}_n(\omega)= \int dt e^{i\omega t} \Re e [C_-(t)C_+(t)]=
 $$
 $$
 =\eta  q(x_n,\omega)\frac{1-\omega/\omega_q}{(1-\omega/\omega_q)^2+\eta^2 | q(x_n,\omega)|^2},
$$
where we introduce the dimensionless strength of interaction, $\eta=\alpha  [\hbar \omega_p/(2E_J)]^2$ and the dimensionless charge distribution, $q(x,t)=Q(x,t)/e$. In the low-dissipative regime we introduce the relaxation time $T$, and by solving the dynamic equations for the density matrix, the time-dependent correlation function of $n$-th qubit is written in the following form:
\begin{equation} \label{CoorFunction}
S_n(\omega)= \eta \frac{1-\omega/\omega_q+i/(\omega_q T)}{(1-\omega/\omega_q)^2+1/(\omega_q T)^2+\eta^2 |q(x_n,\omega)|^2}q(x_n,\omega)
\end{equation}
Substituting (\ref{Average}) and (\ref{CoorFunction}) in (\ref{dynequation}) we obtain the effective equation allowing one to analyze the transmission coefficient $D(\omega)$ for propagating EWs of frequency $\omega$.  
$$
c_0^2 \frac{d^2 q}{d x^2}+\omega^2 q(x) +i\gamma \omega_q(x) =\frac{2 w\hbar \omega_q}{e^2 L_0 \ell} \eta^2 
$$
\begin{equation}
    \label{Propagationequation}
 \sum_{n=1}^N \delta (na-x) \frac{1-\omega/\omega_q+i/(T\omega_q)}{\eta^2 |q(x,\omega)|^2+(1-\omega/\omega_q)^2+1/(T \omega_q)^2}q(x,\omega).
\end{equation}

%Let us suppose that the whole system can be described by function %$\Phi=\begin{bmatrix}\Phi_-\\\Phi_+\end{bmatrix} $. In such a case %Schrödinger equation that describes our system takes the following %form:
%\begin{equation}
%   \ i\begin{bmatrix}\partial_t&0\\0&\ \partial_t\end{bmatrix}\begin{bmatrix}\Phi_-\\\Phi_+\end{bmatrix} =\begin{bmatrix}-\frac{E_J}{2}&-\beta Q\\-\beta Q&\frac{E_J}{2}\end{bmatrix}\begin{bmatrix}\Phi_-\\\Phi_+\end{bmatrix} 
%\end{equation}
%The equation of motion can be written in the following form:
%\begin{equation}
%\delta_i{}_n(\bar{p}_n+\mu_F)=0
%\end{equation}
%Where $mu_F=\omega_p$ and %$\gamma=\omega_C\sqrt[]{\frac{\omega_0}{\omega_L}}$
%the expected value of $\bar{p}_n+\mu_F$ can be found from the %Schrödinger’s equation and takes the following form:
%
\section{EW transmission: a single qubit}
In this Section we consider the EW transmission through a single qubit. The charge distribution $q(x,\omega)$ satisfies the effective equation
$$
c_0^2 \frac{d^2 q}{d x^2}+[\omega^2  +i\gamma \omega ] q(x) =\frac{2\hbar w\omega_q}{e^2 L_0 \ell} \eta^2
$$
\begin{equation}
    \label{Propagationequation-SQ}
\delta (x) \frac{1-\omega/\omega_q+i/(T\omega_q)}{\eta^2 |q(x,\omega)|^2+(1-\omega/\omega_q)^2+1/(T \omega_q)^2}q(x,\omega).
\end{equation}
As the power of EW is small, i.e. $|Q(x)/e|<<(\eta \omega_q T)^{-1}$, the transmission coefficient $D(\omega)$ reads as
\begin{equation}
    \label{TQ-SQ-LC}
    \ D(\omega) =\left[ 1+ \frac{g}{4}\frac{g+4\Gamma}{(\omega/\omega_q-1)^2+\Gamma^2} \right]^{-1}.
\end{equation}
Here, we introduce the dimensionless relaxation rate of a single qubit, $\Gamma=(T\omega_q)^{-1}$ and the interaction strength $g=2\eta^2  \frac{\hbar w}{e^2 c_0  L_0 \ell} $, where $w$  is the geometrical size of the lumped quantum circuit (qubit), $w \ll \ell$. The dependencies of $D(\omega)$ in the linear regime for different values of  $\Gamma$ and $g$ are presented in Fig. 2. A most important effect is a strong resonant suppression of EW propagation in the limit of $g/\Gamma \gg 1$. The width of the $D(\omega)$ curve is diminished as the relaxation rate $\Gamma$ decreases. 

As we turn to the high power regime of applied microwave radiation, i.e. $|q(-\infty)| \simeq \sqrt{P_0}>> \Gamma/\eta $ we obtain the transmission coefficient as a solution of transcendent equation:
\begin{widetext}
\begin{equation}
    \label{TC-SQ-NLR}
D(\omega)  = \left \{ 1+\frac{g}{4}\frac{(4\Gamma +g )[(\omega/\omega_q-1)^2+\Gamma^2] +4\Gamma \eta^2 P_0 D(\omega) }{\left[(\omega/\omega_q-1)^2+\eta^2 P_0 D(\omega)+\Gamma^2 \right ]^2}\right \}^{-1}
\end{equation}
\end{widetext}
Here, $P_0$ is the power of applied microwave radiation far away from the resonator.  An analysis of the Eq. (\ref{TC-SQ-NLR}) shows that in the nonlinear regime the transmission coefficient $D$ is determined strongly by the ratio of two parameters $g$ and $\sqrt{(\omega-\omega_q)^2+\Gamma^2}$. Indeed, if $g/ \sqrt{(\omega-\omega_q)^2+\Gamma^2} \leq 1$ the transmission coefficient just monotonically increases with $P_0$ but in the opposite case, $g/\sqrt{(\omega-\omega_q)^2+\Gamma^2} \gg 1$, there is a particular range of power $P_0$, where two dynamic states of EWs characterized by large and small transmission coefficients, are obtained. The numerically calculated dependencies of the transmission coefficient on the power $P_0$  are shown in Fig. 3.
%    Figure 2
\begin{figure}
\includegraphics[width=1.1\columnwidth]{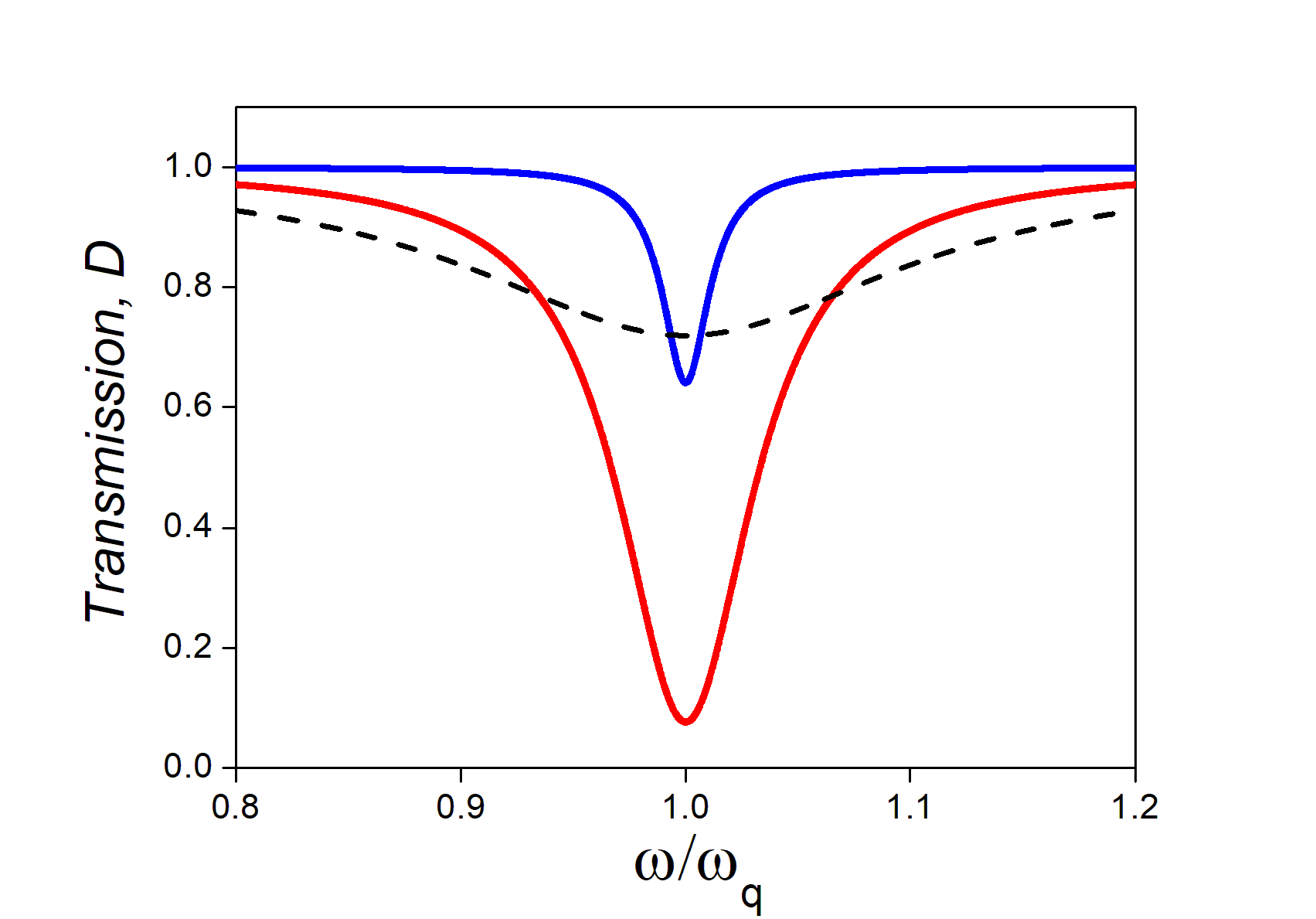}
    \caption{(color online) The transmission of EWs, $D(\omega)$: the  linear regime, a single qubit embedded in the transmission line. The parameters were chosen as: $\Gamma=10^{-2}, g=0.06$ (red solid line), $\Gamma=10^{-2}, g=0.008$ (blue solid line), $\Gamma=10^{-1}, g=0.06$ (black dashed line).}
    \label{fig:2}
\end{figure}
%    Figure 3
\begin{figure}
    \includegraphics[width=1.1\columnwidth]{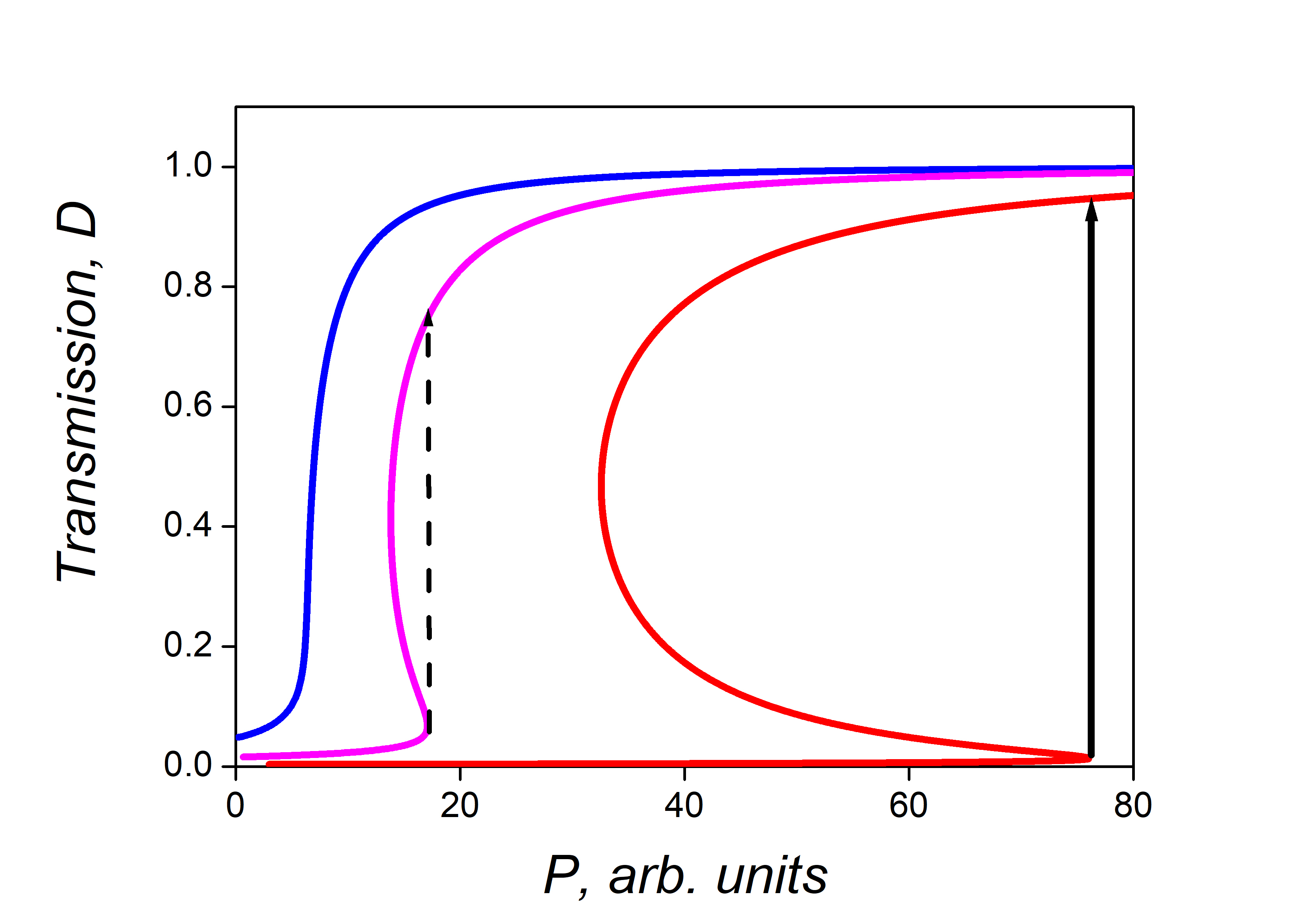}
    \caption{(color online) The transmission coefficient $D$ on  the power $P_0$ of applied microwave radiation. The parameters were chosen as $\omega=\omega_q$ and   $g/\Gamma=9$ (blue line), $g/\Gamma=16$ (magenta line), $g/\Gamma=34.6$ (red line). }
    \label{fig:3}
\end{figure}

%    Figure 3
%\begin{figure}
    %\includegraphics[width=0.75\columnwidth]{graph3%.png}
   % \caption{$\theta=0.1$  $\delta=0.1 (green line) %$ ,$\delta=0.04 (yellow line) $ ,$\delta=0.01 %(red line) $, $\omega=E_j$}
    %\label{fig:schematic}
%\end{figure}

%    Figure 4
%\begin{figure}
 %   \includegraphics[width=0.75\columnwidth]{graph4%.png}
  %  \caption{$\theta=0.1$  $\delta=0.1$ a=1 (green %line),  a=0.9 (yellow line), a=0.01 (red line)}
  %  \label{fig:schematic}
%\end{figure}

%    Figure 5
%\begin{figure}
    %\includegraphics[width=0.75\columnwidth]{graph5%.png}
 % \caption{{$\theta=0.1$  $\delta=0.01$ a=1 (green %line),  a=0.9 (yellow line), a=0.01 (red line)}}
   % \label{fig:schematic}
%\end{figure}
 
\section{EW Transmission: a periodic array of superconducting qubits}
In this Section we consider the propagation of EWs through a periodic array of  $N$ qubits. In this case the charge distribution $q(x,t)$ is determined by  Eq. (\ref{Propagationequation}).  

By making use of the method elaborated for the solution of the Schr\"odinger equation with the Kronig-Penney potential \cite{lifshitz1988introduction} we present the charge distribution $q(x)$ in the following form:
$$
q(x)=-\frac{i  }{2 k c_0^2} \sum_n \beta\{q_n\} q_n \exp{(ik|x-x_n|)}~,
$$
 \begin{equation} \label{Presentation-Q}
\beta=\eta^2 \frac{2\hbar}{e^2} \frac{w \omega_q}{ L_0 \ell} 
\frac{1-\omega/\omega_q+i/(T\omega_q)}{(1-\omega/\omega_q)^2+1/(T \omega_q)^2+\eta^2 |q(x,\omega)|^2}.
\end{equation}
where the wave vector $k=\sqrt{\omega^2+i\gamma \omega}/c_0$, and $q_n=q(x_n)$ is the amplitude of propagating charge distribution at the point of $x_n$. By making use of the properties of the $\delta-$ function, we obtain the set of discrete equations for $q_n$
 \begin{equation} \label{DiscreteEquation}
q_{n+1}+q_{n-1}-\left[ 2\cos ka-\frac{\beta\{q_n\}}{k c_0^2}\sin(ka)\right] q_n=0~.
\end{equation}
Here, $a=\ell/N$ is the  distance between the adjacent qubits in the array. 

Next, we study the EW propagation through an array of densely arranged qubits, i.e. as the condition $ka \ll 1$ is valid. In this case one can transform the difference equation (\ref{DiscreteEquation}) into the differential equation 
 \begin{equation} \label{GenNonlEquation}
c_o^2 \frac{d^2 q(x)}{dx^2}+\left[\omega^2+i\gamma \omega  +\frac{\beta\{q(x)\}}{a} \right] q(x)=0~.
\end{equation}
The  transmission coefficient is determined as $D(\omega)=|q(\ell)/q(0)|^2$.  
%At first we consider the linear regime of a weak %applied radiation. In this case the characteristic %frequency $\Omega$ does not depend on $Q$ and we %obtain the transmission coefficient as

\subsection{Low power regime}

As the power of applied microwave radiation is low, one can neglect the nonlinear dependence of $\beta (q)$ on $q$, and by making use of a well known result for the quantum tunneling through a rectangular barrier the transmission coefficient is written as
%Considering that $\frac{L}{\lambda} %\sqrt[]{\frac{\alpha}{E_J^2 T}} << 1$ 
% we can see that the transmission coefficient takes %the following form 
%$\ D=(1-\alpha \frac{\omega %l}{v}(\frac{\frac{1}{T}+\frac{\omega l}{v} %(E_j-\omega)}{(E_j-\omega)^2+\frac{1}{T^2}}+(\alpha %\frac{\omega l}{v})^2\frac{1+(\frac{\omega %l}{v})^2}{(E_j-\omega)^2+\frac{1}{T^2}})^(-1/2) $

%Let us introduce new parameters, namely, %$p=\frac{\alpha}{E_J}$ which describes the force of %coupling, $q=\frac{E_JL}{v}$ describes energetic %characteristics of the field and Josephson's %elements. 

%    Figure 6.3
%\begin{figure}  
  %   \includegraphics[width=0.75\columnwidth]{graph6_3.%png}
  %  \caption{p=10, q=0.01,$ \delta=0.01$ (red %line),$\delta=0.1$ (blue line)} 
 %   \label{fig:xi1TAlpha}
%\end{figure}
%    Figure 6.4
%\begin{figure}
 %     \includegraphics[width=0.75\columnwidth]{graph6_4%.png}
 %   \caption{p=100, q=0.01,$ \delta=0.01$ (red %line),$\delta=0.1$ (blue line)} 
 %   \label{fig:xi1TAlpha}
%\end{figure}

\begin{equation}
    \label{TrCoeff-Array-Lin}
  \ D = \left | cos(k \ell \sqrt{K (\omega)})+\frac{i}{2}\sqrt{ K(\omega)} sin(k \ell \sqrt{ K(\omega)}) \right | ^{-2}
\end{equation} 
 where $K(\omega)=\frac{g c_0}{\omega_q a} \frac{(1-\omega/\omega_q)+i \Gamma}{(1-\omega/\omega_q)^2+\Gamma^2}$.
 
\begin{figure}
\includegraphics[width=1\columnwidth]{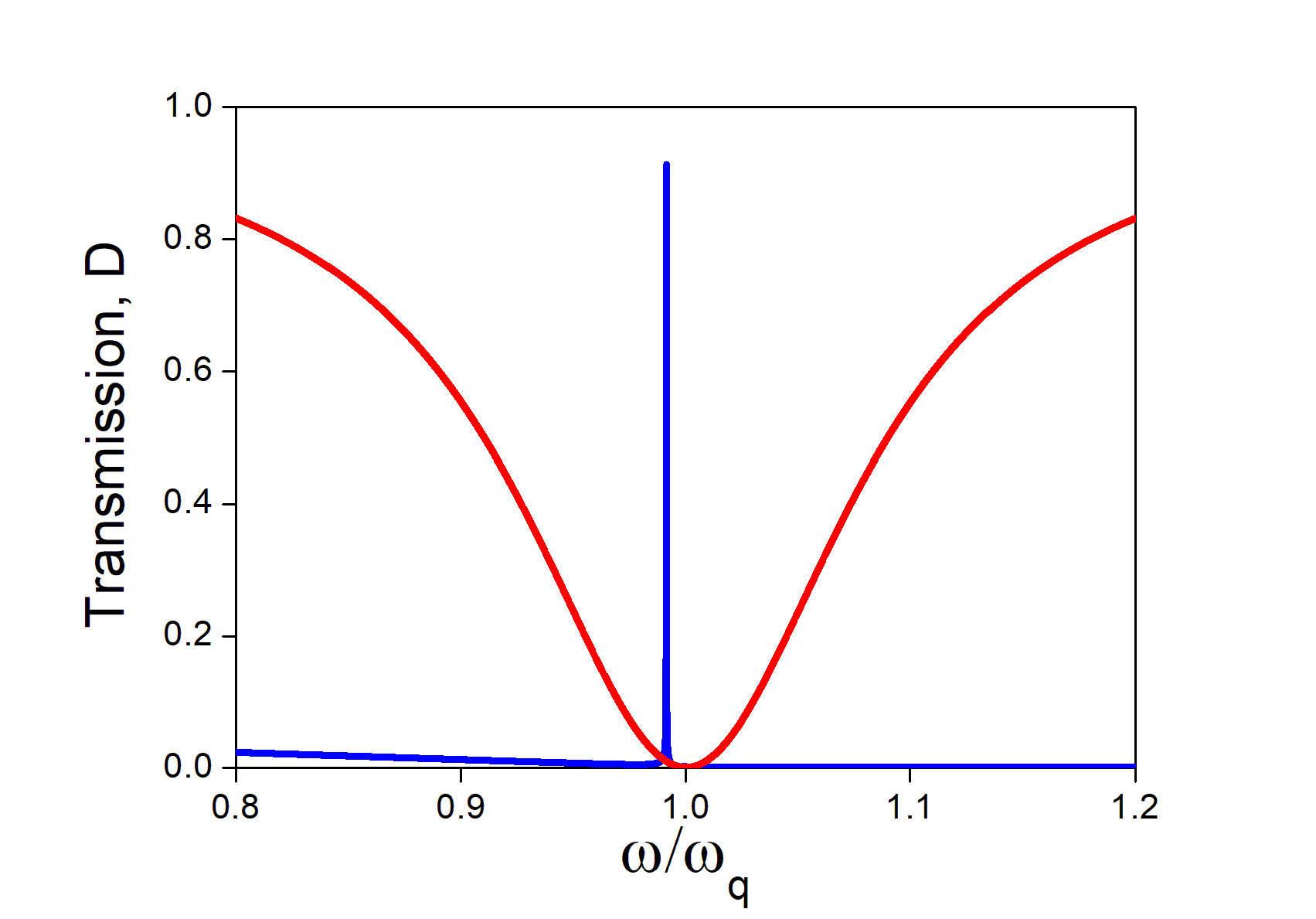}
\caption{(color online) The transmission coefficient of EWs, $D(\omega)$: the  linear regime, a moderate size ($k \ell =0.01$) array of qubits embedded in a low-dissipative transmission line. The parameters were chosen as: $\Gamma=3 \cdot 10^{-3}$ and  $g c_0/(\omega_q a)=9$ (red solid line), $ g c_0/(\omega_q a)=900$ (blue solid line). } 
%g c_0/(\omega_q a)=900$ (black dashed line). 
\label{fig:xi1TAlpha}
\end{figure}

\begin{figure}
\includegraphics[width=1\columnwidth]{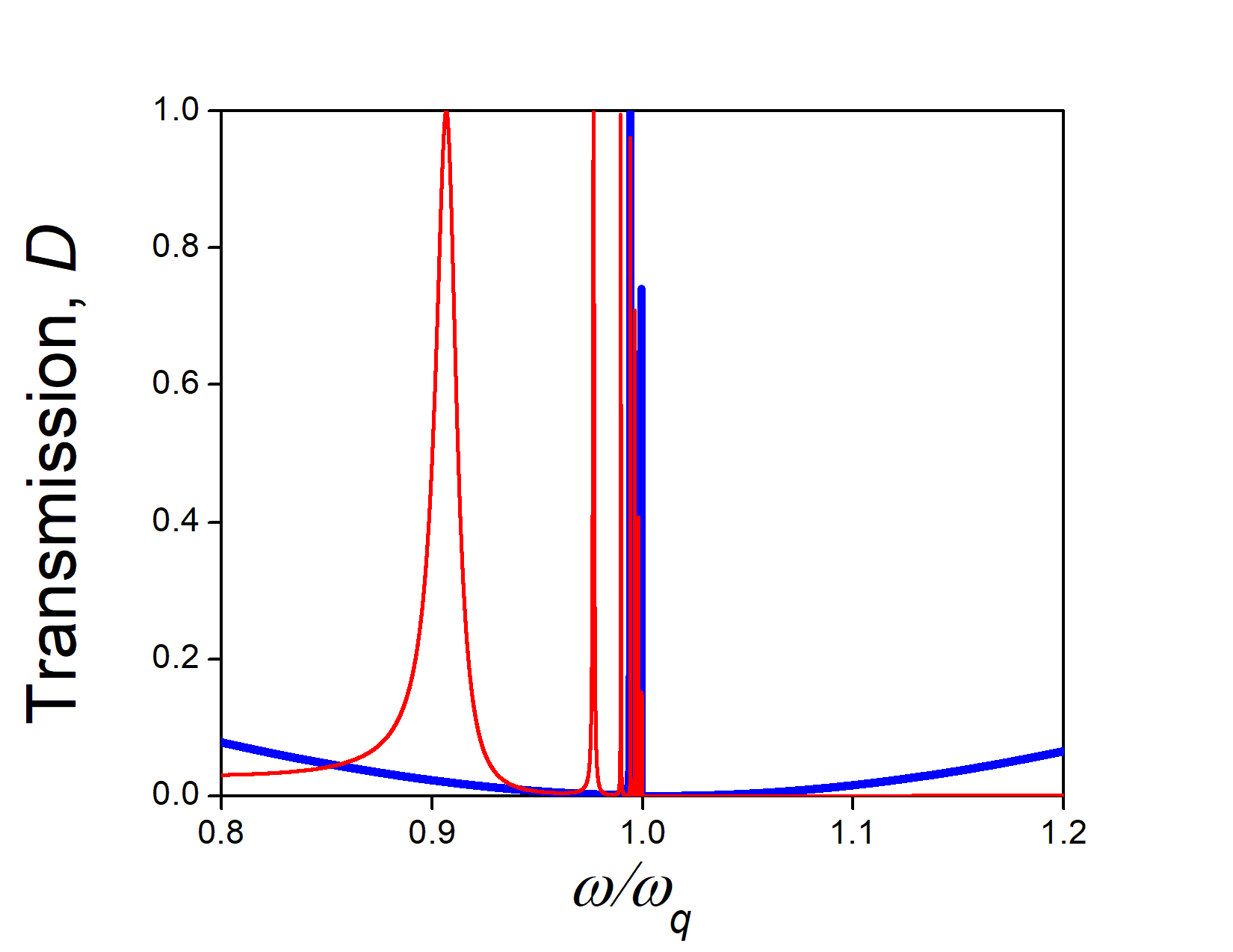}
\caption{(color online) The transmission coefficient of EWs, $D(\omega)$: the  linear regime, large size  arrays of qubits embedded in a low-dissipative transmission line. The parameters were chosen as: $\Gamma=3 \cdot 10^{-3}$,  $g c_0/(\omega_q a)=9$, and different values of $k\ell=0.08$ (blue thick   line), $ k \ell=0.32$ (red thin line). } 
%g c_0/(\omega_q a)=900$ (black dashed line). 
\label{Fig5}
\end{figure}

 Here, we consider a regime most relevant to current experiments  as the total length of a system is smaller than the wave length of EWs, i.e. $\ell \ll \lambda=c_0/\omega$, and the effective strength of interaction between a single qubit and EWs is large, i.e. $\beta/a \gg \omega_q^2$.
 With such assumptions the dependencies of $D(\omega)$ for different values of an effective strength of interaction $g$ are presented in Fig. \ref{fig:xi1TAlpha}. Beyond a standard resonant suppression of $D(\omega)$ observed for moderately large values of $g$ (see Fig.\ref{fig:xi1TAlpha}, red line) , we obtain a great enhancement of $D(\omega)$ in an extremely narrow region of frequencies (see Fig.\ref{fig:xi1TAlpha}, blue line). This effect of resonant transparency of EWs propagating through an array of qubits occurs for an extremely large values of an  effective strength of interaction $g$. We notice that such resonant propagation of EWs through a chain of densely arranged superconducting qubits has been experimentally observed in \cite{shulga2018magnetically} where an extremely large value of coupling was achieved by direct incorporation of qubits Josephson junctions  in the superconducting transmission line. Moreover, as the size of the array increases we obtain a large set of peaks in the dependence of $D(\omega)$. It is shown in Fig. \ref{Fig5}.

%    Figure 7
%\begin{figure}
   % \includegraphics[width=0.75\columnwidth]{graph7_1.p%ng}
  %  \includegraphics[width=0.75\columnwidth]{graph7_2.p%ng}
  %   \includegraphics[width=0.75\columnwidth]{graph7_3.png}
     % \includegraphics[width=0.75\columnwidth]{graph7_4%.png}
   % \caption{} 
  %  \label{fig:xi1TAlpha}
%\end{figure}

%    Figure 8
%\begin{figure}
 %   \includegraphics[width=0.75\columnwidth]{graph8_b.p%ng}
 %   \includegraphics[width=0.75\columnwidth]{graph8_e.p%ng}
 %    \includegraphics[width=0.75\columnwidth]{graph8_f.%png}
 %     \includegraphics[width=0.75\columnwidth]{graph8_j%.png}
 %   \caption{} 
  %  \label{fig:xi1TAlpha}
%\end{figure}

%    Figure 9
%\begin{figure}
 %   \includegraphics[width=0.75\columnwidth]{graph9.png%}
  %  \caption{} 
  %  \label{fig:xi1TAlpha}
%\end{figure}

%    Figure 10
%\begin{figure}
   % \includegraphics[width=0.75\columnwidth]{graph10.pn%g}
   % \caption{} 
   % \label{fig:xi1TAlpha}
%\end{figure}

\subsection{High power regime}

\begin{figure}
    \includegraphics[width=1\columnwidth]{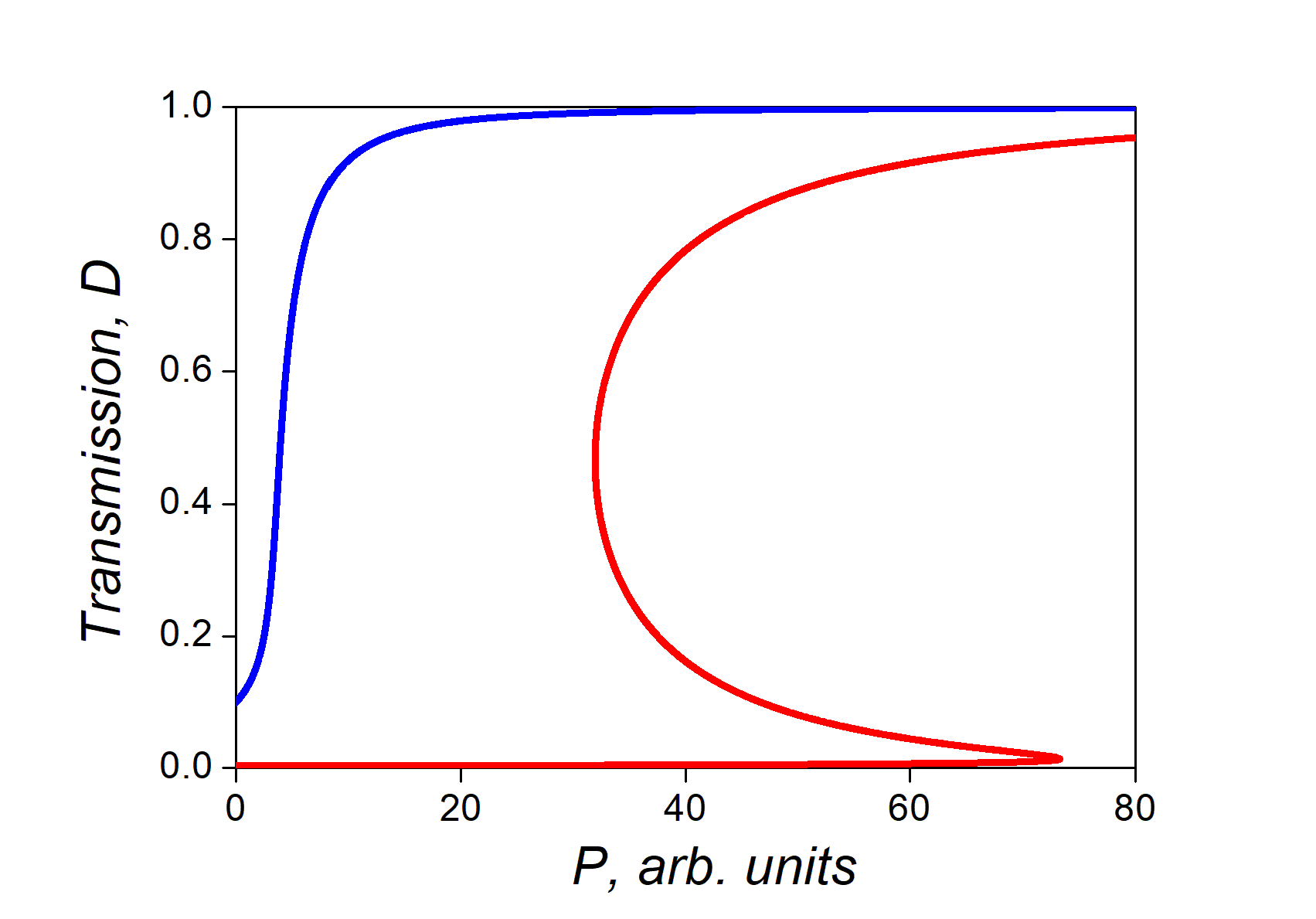}
    \caption{(color online) The EWs transmission through an array of qubits: high power regime. The different values of parameter $\chi \ell/(4k\xi^2)$ are chosen as $4$ (blue line), $17$ (red line).} 
   \label{fig:xi1TAlpha-LP}
\end{figure}

In the regime of high power applied microwave radiation the dynamics of EWs is determined by generic Eq. (\ref{GenNonlEquation})  written as
\begin{equation}
    \label{NonlEq-Arrays}
    \frac{d^2 q}{dx^2}+\left [k^2+\frac{\chi}{|q|^2+\xi^2} \right]q(x)=0,
\end{equation} 
where $\chi=[g c_0/(\eta^2 \omega_q a)](\omega_q-\omega)$  and $\xi=(1/\eta^2)[(\omega_q-\omega)^2+\Gamma^2]$. Here, we neglect a small absorption of EWs, i.e. an imaginary part of $k$ and $\chi$. We  solve such intrinsically nonlinear wave equation by making use of an analogy with the famous  Kepler problem in classical mechanics \cite{landau1960course}. To make that we introduce the spatially dependent amplitude $r(x)$ and the phase $\phi(x)$ of EWs as $q(x)=r(x)e^{i\phi(x)}$ and $|q|=r$. Spatial distributions of the amplitude  $r(x)$ and phase $\phi(x)$ of the electromagnetic field  are determined by  following equations:
$$
r^2 \frac{d\phi}{dx}=C
$$
\begin{equation}
    \label{NonlEquation-ArrQubit}
\frac{d}{dx} \left [(r^\prime)^2+\frac{C^2}{r^2} \right]+(r^2)^\prime \left [k^2+R(r) \right]=0,
\end{equation} 
where we introduce the nonlinear function $R(r)=\frac{\chi}{r^2+\xi^2}$, $C$ is the constant that have to be found from the boundary conditions. The boundary conditions are derived from the continuity of the electric and magnetic  fields of EW at the boundaries, $x=0$ and $x=\ell$, of a system. The boundary conditions are explicitly written down as
\begin{equation}
    \label{BoundaryConditions}
    \begin{cases}
    \frac{d}{dx}\ln q(\ell)=ik\\
   A+B=q(0) \\
    A-B=q^\prime(0)/ik, \\
 \end{cases}
\end{equation}  
where  the amplitude of incident EW, $A \propto \sqrt{P}$ and $P$ is the  power of an incident EW. The transmission coefficient of propagating EWs is determined as $D=|q(\ell)/A|^2$. The solution of Eq. (\ref{NonlEquation-ArrQubit}) is obtained as
\begin{equation}
    \label{SolutionNonlEquation-ArrQubit}
    \int\limits_{r(0)}^{r(\ell)}\frac{du}{  \sqrt{E-\frac{C^2}{r^2}-r^2k^2-\chi \ln(r^2+\xi^2) } }=\ell,
\end{equation}
where the constant $E$ is the effective energy of a system. The constant $C$ is determined as: $C=k[r(\ell)]^2$. In the limit of not extremely large coupling $g$ and large system size, $kl \gg 1$, using the condition $|r(\ell)-r(0)| \ll r(0)$ we write down the expression for the transmission coefficient $D(\omega)$ as 
%solution of Eq. %(\ref{SolutionNonlEquation-ArrQ%ubit}) as
\begin{equation}
    \label{Transmission-Array}
    D=\frac{1}{1+\frac{\chi}{2k^2\xi^2}(1-z)},
\end{equation} 
where the variable $z=r(0)/r(\ell)$ is closed to one.  The parameter $z$ is determined by  the transcendent equation derived from Eq. (\ref{SolutionNonlEquation-ArrQubit}) as
\begin{equation}
    \label{ZParameter}
   \sqrt{\frac{[r^2(\ell)+\xi^2]}{2\chi}}\int\limits_z^1\frac{dy}{\sqrt{1-y}}=\ell
\end{equation} 
Thus, the parameter $z$ is obtained explicitly as 
\begin{equation} \label{ZParameter-2}
    1-z=\frac{\chi \ell^2}{2[r^2(\ell)+\xi^2]}.
\end{equation} 
Substituting  (\ref{ZParameter-2}) in (\ref{Transmission-Array}) we obtain in a strongly nonlinear regime the transmission  coefficient of $D(\omega)$ as
\begin{equation}
   \frac{1}{D}=1+\left [ \frac{\chi \ell}{2k \xi^2 (PD+1)} \right ]^2.
\end{equation} 
For various parameters $\omega_q-\omega$, $\chi$ and $\Gamma$ the dependencies of $D(P)$ are presented in Fig. \ref{fig:xi1TAlpha-LP}.
%    Figure 6.2
Thus, the  main result of this Section is that if for low power EWs the transmission $D$ is strongly suppressed ($D \ll 1$) but in the high-power limit the transmission will be recovered to $D \simeq 1$. The origin of such effect is an equalizing of the populations of qubits states in the limit of a large power of EWs, that,  in turn, strongly suppresses the ac response of qubits to the applied electromagnetic field.  

\section{Conclusion}
In conclusion we theoretically studied the propagation of EWs through a one-dimensional array of densely arranged superconducting qubits, i.e. coherent two-level systems embedded in a low-dissipative transmission line (see Fig. 1). A particular near-resonant case as $\omega \simeq \omega_q$ has been studied. We derive an effective nonlinear wave equation taking into account a non-equilibrium state of qubits, i.e. Eq. (\ref{Propagationequation}). 

The dependencies of transmission coefficient $D(\omega, P)$ on the frequency $\omega$ and power $P$ of applied microwave radiation were obtained. In particular, for both cases of a single qubit and large arrays of qubits the resonant suppression of $D(\omega)$ was found in the limit of small power $P$ and as $|\omega-\omega_q| \ll \omega_q$ (see Fig. 2 and 4). However, the resonant transmission with $D \simeq 1$ was found in large arrays of qubits for an extremely large coupling of qubits with EWs (see Fig. 4 and 5) and in a narrow band of frequencies. Notice here that the effect of resonant transmission of EWs through an array of qubits has been observed in Ref. \cite{shulga2018magnetically}. 

In the limit of high powers of  applied EWs the large transmission $D \simeq 1$ was recovered in both cases of a single qubit and an array of qubits. 

%considered propagation of EWs through cavity coupled to %qubits. 
We anticipate that  strong variations of transmission coefficient $D$ on the frequency and power of EWs will be used in electronic devices with quantum efficiency. 

\begin{acknowledgments}
The authors thank S. Mukhin and S. Flach for useful discussions.  The authors acknowledge a partial financial support of Ministry of Science and Higher Education of the Russian Federation in the framework of Increase Competitiveness Program of NUST 'MISiS' K2-2017-085 and the State Program 3.3360.2017.
\end{acknowledgments}

%\bibliography{general,flatband,josephson}

\begin{thebibliography}{25}%
\makeatletter
\providecommand \@ifxundefined [1]{%
 \@ifx{#1\undefined}
}%
\providecommand \@ifnum [1]{%
 \ifnum #1\expandafter \@firstoftwo
 \else \expandafter \@secondoftwo
 \fi
}%
\providecommand \@ifx [1]{%
 \ifx #1\expandafter \@firstoftwo
 \else \expandafter \@secondoftwo
 \fi
}%
\providecommand \natexlab [1]{#1}%
\providecommand \enquote  [1]{``#1''}%
\providecommand \bibnamefont  [1]{#1}%
\providecommand \bibfnamefont [1]{#1}%
\providecommand \citenamefont [1]{#1}%
\providecommand \href@noop [0]{\@secondoftwo}%
\providecommand \href [0]{\begingroup \@sanitize@url \@href}%
\providecommand \@href[1]{\@@startlink{#1}\@@href}%
\providecommand \@@href[1]{\endgroup#1\@@endlink}%
\providecommand \@sanitize@url [0]{\catcode `\\12\catcode `\$12\catcode
  `\&12\catcode `\#12\catcode `\^12\catcode `\_12\catcode `\%12\relax}%
\providecommand \@@startlink[1]{}%
\providecommand \@@endlink[0]{}%
\providecommand \url  [0]{\begingroup\@sanitize@url \@url }%
\providecommand \@url [1]{\endgroup\@href {#1}{\urlprefix }}%
\providecommand \urlprefix  [0]{URL }%
\providecommand \Eprint [0]{\href }%
\providecommand \doibase [0]{http://dx.doi.org/}%
\providecommand \selectlanguage [0]{\@gobble}%
\providecommand \bibinfo  [0]{\@secondoftwo}%
\providecommand \bibfield  [0]{\@secondoftwo}%
\providecommand \translation [1]{[#1]}%
\providecommand \BibitemOpen [0]{}%
\providecommand \bibitemStop [0]{}%
\providecommand \bibitemNoStop [0]{.\EOS\space}%
\providecommand \EOS [0]{\spacefactor3000\relax}%
\providecommand \BibitemShut  [1]{\csname bibitem#1\endcsname}%
\let\auto@bib@innerbib\@empty
%</preamble>
\bibitem [{\citenamefont {Liao}\ \emph {et~al.}(2016)\citenamefont {Liao},
  \citenamefont {Liu}, \citenamefont {Ma}, \citenamefont {Li}, \citenamefont
  {Jin},\ and\ \citenamefont {Cui}}]{liao2016electromagnetically}%
  \BibitemOpen
  \bibfield  {author} {\bibinfo {author} {\bibfnamefont {Zhen}\ \bibnamefont
  {Liao}}, \bibinfo {author} {\bibfnamefont {Shuo}\ \bibnamefont {Liu}},
  \bibinfo {author} {\bibfnamefont {Hui~Feng}\ \bibnamefont {Ma}}, \bibinfo
  {author} {\bibfnamefont {Chun}\ \bibnamefont {Li}}, \bibinfo {author}
  {\bibfnamefont {Biaobing}\ \bibnamefont {Jin}}, \ and\ \bibinfo {author}
  {\bibfnamefont {Tie~Jun}\ \bibnamefont {Cui}},\ }\bibfield  {title} {\enquote
  {\bibinfo {title} {Electromagnetically induced transparency metamaterial
  based on spoof localized surface plasmons at terahertz frequencies},}\
  }\href@noop {} {\bibfield  {journal} {\bibinfo  {journal} {Scientific
  reports}\ }\textbf {\bibinfo {volume} {6}},\ \bibinfo {pages} {27596}
  (\bibinfo {year} {2016})}\BibitemShut {NoStop}%
\bibitem [{\citenamefont {Shulga}\ \emph {et~al.}(2018)\citenamefont {Shulga},
  \citenamefont {Il’ichev}, \citenamefont {Fistul}, \citenamefont {Besedin},
  \citenamefont {Butz}, \citenamefont {Astafiev}, \citenamefont {H{\"u}bner},\
  and\ \citenamefont {Ustinov}}]{shulga2018magnetically}%
  \BibitemOpen
  \bibfield  {author} {\bibinfo {author} {\bibfnamefont {KV}~\bibnamefont
  {Shulga}}, \bibinfo {author} {\bibfnamefont {E}~\bibnamefont {Il’ichev}},
  \bibinfo {author} {\bibfnamefont {MV}~\bibnamefont {Fistul}}, \bibinfo
  {author} {\bibfnamefont {IS}~\bibnamefont {Besedin}}, \bibinfo {author}
  {\bibfnamefont {S}~\bibnamefont {Butz}}, \bibinfo {author} {\bibfnamefont
  {OV}~\bibnamefont {Astafiev}}, \bibinfo {author} {\bibfnamefont
  {U}~\bibnamefont {H{\"u}bner}}, \ and\ \bibinfo {author} {\bibfnamefont
  {AV}~\bibnamefont {Ustinov}},\ }\bibfield  {title} {\enquote {\bibinfo
  {title} {Magnetically induced transparency of a quantum metamaterial composed
  of twin flux qubits},}\ }\href@noop {} {\bibfield  {journal} {\bibinfo
  {journal} {Nature communications}\ }\textbf {\bibinfo {volume} {9}},\
  \bibinfo {pages} {150} (\bibinfo {year} {2018})}\BibitemShut {NoStop}%
\bibitem [{\citenamefont {Chaldyshev}\ \emph {et~al.}(2011)\citenamefont
  {Chaldyshev}, \citenamefont {Bolshakov}, \citenamefont {Zavarin},
  \citenamefont {Sakharov}, \citenamefont {Lundin}, \citenamefont
  {Tsatsulnikov}, \citenamefont {Yagovkina}, \citenamefont {Kim},\ and\
  \citenamefont {Park}}]{chaldyshev2011optical}%
  \BibitemOpen
  \bibfield  {author} {\bibinfo {author} {\bibfnamefont {VV}~\bibnamefont
  {Chaldyshev}}, \bibinfo {author} {\bibfnamefont {AS}~\bibnamefont
  {Bolshakov}}, \bibinfo {author} {\bibfnamefont {EE}~\bibnamefont {Zavarin}},
  \bibinfo {author} {\bibfnamefont {AV}~\bibnamefont {Sakharov}}, \bibinfo
  {author} {\bibfnamefont {WV}~\bibnamefont {Lundin}}, \bibinfo {author}
  {\bibfnamefont {AF}~\bibnamefont {Tsatsulnikov}}, \bibinfo {author}
  {\bibfnamefont {MA}~\bibnamefont {Yagovkina}}, \bibinfo {author}
  {\bibfnamefont {Taek}\ \bibnamefont {Kim}}, \ and\ \bibinfo {author}
  {\bibfnamefont {Youngsoo}\ \bibnamefont {Park}},\ }\bibfield  {title}
  {\enquote {\bibinfo {title} {Optical lattices of ingan quantum well
  excitons},}\ }\href@noop {} {\bibfield  {journal} {\bibinfo  {journal}
  {Applied Physics Letters}\ }\textbf {\bibinfo {volume} {99}},\ \bibinfo
  {pages} {251103} (\bibinfo {year} {2011})}\BibitemShut {NoStop}%
\bibitem [{\citenamefont {Smith}\ \emph {et~al.}(2004)\citenamefont {Smith},
  \citenamefont {Pendry},\ and\ \citenamefont
  {Wiltshire}}]{smith2004metamaterials}%
  \BibitemOpen
  \bibfield  {author} {\bibinfo {author} {\bibfnamefont {David~R}\ \bibnamefont
  {Smith}}, \bibinfo {author} {\bibfnamefont {John~B}\ \bibnamefont {Pendry}},
  \ and\ \bibinfo {author} {\bibfnamefont {Mike~CK}\ \bibnamefont
  {Wiltshire}},\ }\bibfield  {title} {\enquote {\bibinfo {title} {Metamaterials
  and negative refractive index},}\ }\href@noop {} {\bibfield  {journal}
  {\bibinfo  {journal} {Science}\ }\textbf {\bibinfo {volume} {305}},\ \bibinfo
  {pages} {788--792} (\bibinfo {year} {2004})}\BibitemShut {NoStop}%
\bibitem [{\citenamefont {Zharov}\ \emph {et~al.}(2003)\citenamefont {Zharov},
  \citenamefont {Shadrivov},\ and\ \citenamefont
  {Kivshar}}]{zharov2003nonlinear}%
  \BibitemOpen
  \bibfield  {author} {\bibinfo {author} {\bibfnamefont {Alexander~A}\
  \bibnamefont {Zharov}}, \bibinfo {author} {\bibfnamefont {Ilya~V}\
  \bibnamefont {Shadrivov}}, \ and\ \bibinfo {author} {\bibfnamefont {Yuri~S}\
  \bibnamefont {Kivshar}},\ }\bibfield  {title} {\enquote {\bibinfo {title}
  {Nonlinear properties of left-handed metamaterials},}\ }\href@noop {}
  {\bibfield  {journal} {\bibinfo  {journal} {Physical Review Letters}\
  }\textbf {\bibinfo {volume} {91}},\ \bibinfo {pages} {037401} (\bibinfo
  {year} {2003})}\BibitemShut {NoStop}%
\bibitem [{\citenamefont {Lazarides}\ \emph {et~al.}(2015)\citenamefont
  {Lazarides}, \citenamefont {Neofotistos},\ and\ \citenamefont
  {Tsironis}}]{lazarides2015chimeras}%
  \BibitemOpen
  \bibfield  {author} {\bibinfo {author} {\bibfnamefont {N}~\bibnamefont
  {Lazarides}}, \bibinfo {author} {\bibfnamefont {G}~\bibnamefont
  {Neofotistos}}, \ and\ \bibinfo {author} {\bibfnamefont {GP}~\bibnamefont
  {Tsironis}},\ }\bibfield  {title} {\enquote {\bibinfo {title} {Chimeras in
  squid metamaterials},}\ }\href@noop {} {\bibfield  {journal} {\bibinfo
  {journal} {Physical Review B}\ }\textbf {\bibinfo {volume} {91}},\ \bibinfo
  {pages} {054303} (\bibinfo {year} {2015})}\BibitemShut {NoStop}%
\bibitem [{\citenamefont {Jung}\ \emph {et~al.}(2014)\citenamefont {Jung},
  \citenamefont {Butz}, \citenamefont {Marthaler}, \citenamefont {Fistul},
  \citenamefont {Lepp{\"a}kangas}, \citenamefont {Koshelets},\ and\
  \citenamefont {Ustinov}}]{jung2014multistability}%
  \BibitemOpen
  \bibfield  {author} {\bibinfo {author} {\bibfnamefont {P}~\bibnamefont
  {Jung}}, \bibinfo {author} {\bibfnamefont {S}~\bibnamefont {Butz}}, \bibinfo
  {author} {\bibfnamefont {M}~\bibnamefont {Marthaler}}, \bibinfo {author}
  {\bibfnamefont {MV}~\bibnamefont {Fistul}}, \bibinfo {author} {\bibfnamefont
  {Juha}\ \bibnamefont {Lepp{\"a}kangas}}, \bibinfo {author} {\bibfnamefont
  {VP}~\bibnamefont {Koshelets}}, \ and\ \bibinfo {author} {\bibfnamefont
  {AV}~\bibnamefont {Ustinov}},\ }\bibfield  {title} {\enquote {\bibinfo
  {title} {Multistability and switching in a superconducting metamaterial},}\
  }\href@noop {} {\bibfield  {journal} {\bibinfo  {journal} {Nature
  communications}\ }\textbf {\bibinfo {volume} {5}},\ \bibinfo {pages} {3730}
  (\bibinfo {year} {2014})}\BibitemShut {NoStop}%
\bibitem [{\citenamefont {Ricci}\ \emph {et~al.}(2005)\citenamefont {Ricci},
  \citenamefont {Orloff},\ and\ \citenamefont
  {Anlage}}]{ricci2005superconducting}%
  \BibitemOpen
  \bibfield  {author} {\bibinfo {author} {\bibfnamefont {Michael}\ \bibnamefont
  {Ricci}}, \bibinfo {author} {\bibfnamefont {Nathan}\ \bibnamefont {Orloff}},
  \ and\ \bibinfo {author} {\bibfnamefont {Steven~M}\ \bibnamefont {Anlage}},\
  }\bibfield  {title} {\enquote {\bibinfo {title} {Superconducting
  metamaterials},}\ }\href@noop {} {\bibfield  {journal} {\bibinfo  {journal}
  {Applied Physics Letters}\ }\textbf {\bibinfo {volume} {87}},\ \bibinfo
  {pages} {034102} (\bibinfo {year} {2005})}\BibitemShut {NoStop}%
\bibitem [{\citenamefont {Anlage}(2010)}]{anlage2010physics}%
  \BibitemOpen
  \bibfield  {author} {\bibinfo {author} {\bibfnamefont {Steven~M}\
  \bibnamefont {Anlage}},\ }\bibfield  {title} {\enquote {\bibinfo {title} {The
  physics and applications of superconducting metamaterials},}\ }\href@noop {}
  {\bibfield  {journal} {\bibinfo  {journal} {Journal of Optics}\ }\textbf
  {\bibinfo {volume} {13}},\ \bibinfo {pages} {024001} (\bibinfo {year}
  {2010})}\BibitemShut {NoStop}%
\bibitem [{\citenamefont {Miroshnichenko}\ \emph {et~al.}(2001)\citenamefont
  {Miroshnichenko}, \citenamefont {Flach}, \citenamefont {Fistul},
  \citenamefont {Zolotaryuk},\ and\ \citenamefont
  {Page}}]{miroshnichenko2001breathers}%
  \BibitemOpen
  \bibfield  {author} {\bibinfo {author} {\bibfnamefont {AE}~\bibnamefont
  {Miroshnichenko}}, \bibinfo {author} {\bibfnamefont {S}~\bibnamefont
  {Flach}}, \bibinfo {author} {\bibfnamefont {MV}~\bibnamefont {Fistul}},
  \bibinfo {author} {\bibfnamefont {Yaroslav}\ \bibnamefont {Zolotaryuk}}, \
  and\ \bibinfo {author} {\bibfnamefont {JB}~\bibnamefont {Page}},\ }\bibfield
  {title} {\enquote {\bibinfo {title} {Breathers in josephson junction ladders:
  Resonances and electromagnetic wave spectroscopy},}\ }\href@noop {}
  {\bibfield  {journal} {\bibinfo  {journal} {Physical Review E}\ }\textbf
  {\bibinfo {volume} {64}},\ \bibinfo {pages} {066601} (\bibinfo {year}
  {2001})}\BibitemShut {NoStop}%
\bibitem [{\citenamefont {Filatrella}\ \emph {et~al.}(2000)\citenamefont
  {Filatrella}, \citenamefont {Pedersen},\ and\ \citenamefont
  {Wiesenfeld}}]{filatrella2000high}%
  \BibitemOpen
  \bibfield  {author} {\bibinfo {author} {\bibfnamefont {Giovanni}\
  \bibnamefont {Filatrella}}, \bibinfo {author} {\bibfnamefont {Niels~Falsig}\
  \bibnamefont {Pedersen}}, \ and\ \bibinfo {author} {\bibfnamefont {Kurt}\
  \bibnamefont {Wiesenfeld}},\ }\bibfield  {title} {\enquote {\bibinfo {title}
  {High-q cavity-induced synchronization in oscillator arrays},}\ }\href@noop
  {} {\bibfield  {journal} {\bibinfo  {journal} {Physical Review E}\ }\textbf
  {\bibinfo {volume} {61}},\ \bibinfo {pages} {2513} (\bibinfo {year}
  {2000})}\BibitemShut {NoStop}%
\bibitem [{\citenamefont {Pashkin}\ \emph {et~al.}(2003)\citenamefont
  {Pashkin}, \citenamefont {Yamamoto}, \citenamefont {Astafiev}, \citenamefont
  {Nakamura}, \citenamefont {Averin},\ and\ \citenamefont
  {Tsai}}]{pashkin2003quantum}%
  \BibitemOpen
  \bibfield  {author} {\bibinfo {author} {\bibfnamefont {Yu~A}\ \bibnamefont
  {Pashkin}}, \bibinfo {author} {\bibfnamefont {T}~\bibnamefont {Yamamoto}},
  \bibinfo {author} {\bibfnamefont {O}~\bibnamefont {Astafiev}}, \bibinfo
  {author} {\bibfnamefont {Yasunobu}\ \bibnamefont {Nakamura}}, \bibinfo
  {author} {\bibfnamefont {DV}~\bibnamefont {Averin}}, \ and\ \bibinfo {author}
  {\bibfnamefont {JS}~\bibnamefont {Tsai}},\ }\bibfield  {title} {\enquote
  {\bibinfo {title} {Quantum oscillations in two coupled charge qubits},}\
  }\href@noop {} {\bibfield  {journal} {\bibinfo  {journal} {Nature}\ }\textbf
  {\bibinfo {volume} {421}},\ \bibinfo {pages} {823} (\bibinfo {year}
  {2003})}\BibitemShut {NoStop}%
\bibitem [{\citenamefont {Houck}\ \emph {et~al.}(2009)\citenamefont {Houck},
  \citenamefont {Koch}, \citenamefont {Devoret}, \citenamefont {Girvin},\ and\
  \citenamefont {Schoelkopf}}]{houck2009life}%
  \BibitemOpen
  \bibfield  {author} {\bibinfo {author} {\bibfnamefont {Andrew~A}\
  \bibnamefont {Houck}}, \bibinfo {author} {\bibfnamefont {Jens}\ \bibnamefont
  {Koch}}, \bibinfo {author} {\bibfnamefont {Michel~H}\ \bibnamefont
  {Devoret}}, \bibinfo {author} {\bibfnamefont {Steven~M}\ \bibnamefont
  {Girvin}}, \ and\ \bibinfo {author} {\bibfnamefont {Robert~J}\ \bibnamefont
  {Schoelkopf}},\ }\bibfield  {title} {\enquote {\bibinfo {title} {Life after
  charge noise: recent results with transmon qubits},}\ }\href@noop {}
  {\bibfield  {journal} {\bibinfo  {journal} {Quantum Information Processing}\
  }\textbf {\bibinfo {volume} {8}},\ \bibinfo {pages} {105--115} (\bibinfo
  {year} {2009})}\BibitemShut {NoStop}%
\bibitem [{\citenamefont {Chiorescu}\ \emph {et~al.}(2004)\citenamefont
  {Chiorescu}, \citenamefont {Bertet}, \citenamefont {Semba}, \citenamefont
  {Nakamura}, \citenamefont {Harmans},\ and\ \citenamefont
  {Mooij}}]{chiorescu2004coherent}%
  \BibitemOpen
  \bibfield  {author} {\bibinfo {author} {\bibfnamefont {I}~\bibnamefont
  {Chiorescu}}, \bibinfo {author} {\bibfnamefont {P}~\bibnamefont {Bertet}},
  \bibinfo {author} {\bibfnamefont {K}~\bibnamefont {Semba}}, \bibinfo {author}
  {\bibfnamefont {Y}~\bibnamefont {Nakamura}}, \bibinfo {author} {\bibfnamefont
  {CJPM}\ \bibnamefont {Harmans}}, \ and\ \bibinfo {author} {\bibfnamefont
  {JE}~\bibnamefont {Mooij}},\ }\bibfield  {title} {\enquote {\bibinfo {title}
  {Coherent dynamics of a flux qubit coupled to a harmonic oscillator},}\
  }\href@noop {} {\bibfield  {journal} {\bibinfo  {journal} {Nature}\ }\textbf
  {\bibinfo {volume} {431}},\ \bibinfo {pages} {159} (\bibinfo {year}
  {2004})}\BibitemShut {NoStop}%
\bibitem [{\citenamefont {Majer}\ \emph {et~al.}(2007)\citenamefont {Majer},
  \citenamefont {Chow}, \citenamefont {Gambetta}, \citenamefont {Koch},
  \citenamefont {Johnson}, \citenamefont {Schreier}, \citenamefont {Frunzio},
  \citenamefont {Schuster}, \citenamefont {Houck}, \citenamefont {Wallraff}
  \emph {et~al.}}]{majer2007coupling}%
  \BibitemOpen
  \bibfield  {author} {\bibinfo {author} {\bibfnamefont {J}~\bibnamefont
  {Majer}}, \bibinfo {author} {\bibfnamefont {JM}~\bibnamefont {Chow}},
  \bibinfo {author} {\bibfnamefont {JM}~\bibnamefont {Gambetta}}, \bibinfo
  {author} {\bibfnamefont {Jens}\ \bibnamefont {Koch}}, \bibinfo {author}
  {\bibfnamefont {BR}~\bibnamefont {Johnson}}, \bibinfo {author} {\bibfnamefont
  {JA}~\bibnamefont {Schreier}}, \bibinfo {author} {\bibfnamefont
  {L}~\bibnamefont {Frunzio}}, \bibinfo {author} {\bibfnamefont
  {DI}~\bibnamefont {Schuster}}, \bibinfo {author} {\bibfnamefont
  {AA}~\bibnamefont {Houck}}, \bibinfo {author} {\bibfnamefont {Andreas}\
  \bibnamefont {Wallraff}},  \emph {et~al.},\ }\bibfield  {title} {\enquote
  {\bibinfo {title} {Coupling superconducting qubits via a cavity bus},}\
  }\href@noop {} {\bibfield  {journal} {\bibinfo  {journal} {Nature}\ }\textbf
  {\bibinfo {volume} {449}},\ \bibinfo {pages} {443} (\bibinfo {year}
  {2007})}\BibitemShut {NoStop}%
\bibitem [{\citenamefont {Fink}\ \emph {et~al.}(2009)\citenamefont {Fink},
  \citenamefont {Bianchetti}, \citenamefont {Baur}, \citenamefont {G{\"o}ppl},
  \citenamefont {Steffen}, \citenamefont {Filipp}, \citenamefont {Leek},
  \citenamefont {Blais},\ and\ \citenamefont {Wallraff}}]{fink2009dressed}%
  \BibitemOpen
  \bibfield  {author} {\bibinfo {author} {\bibfnamefont {JM}~\bibnamefont
  {Fink}}, \bibinfo {author} {\bibfnamefont {R}~\bibnamefont {Bianchetti}},
  \bibinfo {author} {\bibfnamefont {Matthias}\ \bibnamefont {Baur}}, \bibinfo
  {author} {\bibfnamefont {M}~\bibnamefont {G{\"o}ppl}}, \bibinfo {author}
  {\bibfnamefont {Lars}\ \bibnamefont {Steffen}}, \bibinfo {author}
  {\bibfnamefont {Stefan}\ \bibnamefont {Filipp}}, \bibinfo {author}
  {\bibfnamefont {PJ}~\bibnamefont {Leek}}, \bibinfo {author} {\bibfnamefont
  {Alexandre}\ \bibnamefont {Blais}}, \ and\ \bibinfo {author} {\bibfnamefont
  {Andreas}\ \bibnamefont {Wallraff}},\ }\bibfield  {title} {\enquote {\bibinfo
  {title} {Dressed collective qubit states and the tavis-cummings model in
  circuit qed},}\ }\href@noop {} {\bibfield  {journal} {\bibinfo  {journal}
  {Physical review letters}\ }\textbf {\bibinfo {volume} {103}},\ \bibinfo
  {pages} {083601} (\bibinfo {year} {2009})}\BibitemShut {NoStop}%
\bibitem [{\citenamefont {Macha}\ \emph {et~al.}(2014)\citenamefont {Macha},
  \citenamefont {Oelsner}, \citenamefont {Reiner}, \citenamefont {Marthaler},
  \citenamefont {Andr{\'e}}, \citenamefont {Sch{\"o}n}, \citenamefont
  {H{\"u}bner}, \citenamefont {Meyer}, \citenamefont {Il’ichev},\ and\
  \citenamefont {Ustinov}}]{macha2014implementation}%
  \BibitemOpen
  \bibfield  {author} {\bibinfo {author} {\bibfnamefont {Pascal}\ \bibnamefont
  {Macha}}, \bibinfo {author} {\bibfnamefont {Gregor}\ \bibnamefont {Oelsner}},
  \bibinfo {author} {\bibfnamefont {Jan-Michael}\ \bibnamefont {Reiner}},
  \bibinfo {author} {\bibfnamefont {Michael}\ \bibnamefont {Marthaler}},
  \bibinfo {author} {\bibfnamefont {Stephan}\ \bibnamefont {Andr{\'e}}},
  \bibinfo {author} {\bibfnamefont {Gerd}\ \bibnamefont {Sch{\"o}n}}, \bibinfo
  {author} {\bibfnamefont {Uwe}\ \bibnamefont {H{\"u}bner}}, \bibinfo {author}
  {\bibfnamefont {Hans-Georg}\ \bibnamefont {Meyer}}, \bibinfo {author}
  {\bibfnamefont {Evgeni}\ \bibnamefont {Il’ichev}}, \ and\ \bibinfo {author}
  {\bibfnamefont {Alexey~V}\ \bibnamefont {Ustinov}},\ }\bibfield  {title}
  {\enquote {\bibinfo {title} {Implementation of a quantum metamaterial using
  superconducting qubits},}\ }\href@noop {} {\bibfield  {journal} {\bibinfo
  {journal} {Nature communications}\ }\textbf {\bibinfo {volume} {5}},\
  \bibinfo {pages} {5146} (\bibinfo {year} {2014})}\BibitemShut {NoStop}%
\bibitem [{\citenamefont {Shulga}\ \emph {et~al.}(2017)\citenamefont {Shulga},
  \citenamefont {Yang}, \citenamefont {Fedorov}, \citenamefont {Fistul},
  \citenamefont {Weides},\ and\ \citenamefont
  {Ustinov}}]{shulga2017observation}%
  \BibitemOpen
  \bibfield  {author} {\bibinfo {author} {\bibfnamefont {Kirill~Vladimirovich}\
  \bibnamefont {Shulga}}, \bibinfo {author} {\bibfnamefont {Ping}\ \bibnamefont
  {Yang}}, \bibinfo {author} {\bibfnamefont {GP}~\bibnamefont {Fedorov}},
  \bibinfo {author} {\bibfnamefont {Mikhail~Viktorovich}\ \bibnamefont
  {Fistul}}, \bibinfo {author} {\bibfnamefont {Martian}\ \bibnamefont
  {Weides}}, \ and\ \bibinfo {author} {\bibfnamefont {Aleksei~Valentinovich}\
  \bibnamefont {Ustinov}},\ }\bibfield  {title} {\enquote {\bibinfo {title}
  {Observation of a collective mode of an array of transmon qubits},}\
  }\href@noop {} {\bibfield  {journal} {\bibinfo  {journal} {JETP letters}\
  }\textbf {\bibinfo {volume} {105}},\ \bibinfo {pages} {47--50} (\bibinfo
  {year} {2017})}\BibitemShut {NoStop}%
\bibitem [{\citenamefont {Fistul}(2017)}]{fistul2017quantum}%
  \BibitemOpen
  \bibfield  {author} {\bibinfo {author} {\bibfnamefont {MV}~\bibnamefont
  {Fistul}},\ }\bibfield  {title} {\enquote {\bibinfo {title} {Quantum
  synchronization in disordered superconducting metamaterials},}\ }\href@noop
  {} {\bibfield  {journal} {\bibinfo  {journal} {Scientific reports}\ }\textbf
  {\bibinfo {volume} {7}},\ \bibinfo {pages} {43657} (\bibinfo {year}
  {2017})}\BibitemShut {NoStop}%
\bibitem [{\citenamefont {Asai}\ \emph {et~al.}(2015)\citenamefont {Asai},
  \citenamefont {Savel'ev}, \citenamefont {Kawabata},\ and\ \citenamefont
  {Zagoskin}}]{asai2015effects}%
  \BibitemOpen
  \bibfield  {author} {\bibinfo {author} {\bibfnamefont {Hidehiro}\
  \bibnamefont {Asai}}, \bibinfo {author} {\bibfnamefont {Sergey}\ \bibnamefont
  {Savel'ev}}, \bibinfo {author} {\bibfnamefont {Shiro}\ \bibnamefont
  {Kawabata}}, \ and\ \bibinfo {author} {\bibfnamefont {Alexandre~M}\
  \bibnamefont {Zagoskin}},\ }\bibfield  {title} {\enquote {\bibinfo {title}
  {Effects of lasing in a one-dimensional quantum metamaterial},}\ }\href@noop
  {} {\bibfield  {journal} {\bibinfo  {journal} {Physical Review B}\ }\textbf
  {\bibinfo {volume} {91}},\ \bibinfo {pages} {134513} (\bibinfo {year}
  {2015})}\BibitemShut {NoStop}%
\bibitem [{\citenamefont {Ivi{\'c}}\ \emph {et~al.}(2016)\citenamefont
  {Ivi{\'c}}, \citenamefont {Lazarides},\ and\ \citenamefont
  {Tsironis}}]{ivic2016qubit}%
  \BibitemOpen
  \bibfield  {author} {\bibinfo {author} {\bibfnamefont {Z}~\bibnamefont
  {Ivi{\'c}}}, \bibinfo {author} {\bibfnamefont {N}~\bibnamefont {Lazarides}},
  \ and\ \bibinfo {author} {\bibfnamefont {GP}~\bibnamefont {Tsironis}},\
  }\bibfield  {title} {\enquote {\bibinfo {title} {Qubit lattice coherence
  induced by electromagnetic pulses in superconducting metamaterials},}\
  }\href@noop {} {\bibfield  {journal} {\bibinfo  {journal} {Scientific
  reports}\ }\textbf {\bibinfo {volume} {6}},\ \bibinfo {pages} {29374}
  (\bibinfo {year} {2016})}\BibitemShut {NoStop}%
\bibitem [{\citenamefont {Iontsev}\ \emph {et~al.}(2016)\citenamefont
  {Iontsev}, \citenamefont {Mukhin},\ and\ \citenamefont
  {Fistul}}]{iontsev2016double}%
  \BibitemOpen
  \bibfield  {author} {\bibinfo {author} {\bibfnamefont {MA}~\bibnamefont
  {Iontsev}}, \bibinfo {author} {\bibfnamefont {SI}~\bibnamefont {Mukhin}}, \
  and\ \bibinfo {author} {\bibfnamefont {MV}~\bibnamefont {Fistul}},\
  }\bibfield  {title} {\enquote {\bibinfo {title} {Double-resonance response of
  a superconducting quantum metamaterial: Manifestation of nonclassical states
  of photons},}\ }\href@noop {} {\bibfield  {journal} {\bibinfo  {journal}
  {Physical Review B}\ }\textbf {\bibinfo {volume} {94}},\ \bibinfo {pages}
  {174510} (\bibinfo {year} {2016})}\BibitemShut {NoStop}%
\bibitem [{\citenamefont {Rakhmanov}\ \emph {et~al.}(2008)\citenamefont
  {Rakhmanov}, \citenamefont {Zagoskin}, \citenamefont {Savel’ev},\ and\
  \citenamefont {Nori}}]{rakhmanov2008quantum}%
  \BibitemOpen
  \bibfield  {author} {\bibinfo {author} {\bibfnamefont {Alexander~L}\
  \bibnamefont {Rakhmanov}}, \bibinfo {author} {\bibfnamefont {Alexandre~M}\
  \bibnamefont {Zagoskin}}, \bibinfo {author} {\bibfnamefont {Sergey}\
  \bibnamefont {Savel’ev}}, \ and\ \bibinfo {author} {\bibfnamefont {Franco}\
  \bibnamefont {Nori}},\ }\bibfield  {title} {\enquote {\bibinfo {title}
  {Quantum metamaterials: Electromagnetic waves in a josephson qubit line},}\
  }\href@noop {} {\bibfield  {journal} {\bibinfo  {journal} {Physical Review
  B}\ }\textbf {\bibinfo {volume} {77}},\ \bibinfo {pages} {144507} (\bibinfo
  {year} {2008})}\BibitemShut {NoStop}%
\bibitem [{\citenamefont {Lifshitz}\ \emph {et~al.}(1988)\citenamefont
  {Lifshitz}, \citenamefont {Gredeskul},\ and\ \citenamefont
  {Pastur}}]{lifshitz1988introduction}%
  \BibitemOpen
  \bibfield  {author} {\bibinfo {author} {\bibfnamefont {IM}~\bibnamefont
  {Lifshitz}}, \bibinfo {author} {\bibfnamefont {SA}~\bibnamefont {Gredeskul}},
  \ and\ \bibinfo {author} {\bibfnamefont {LA}~\bibnamefont {Pastur}},\
  }\bibfield  {title} {\enquote {\bibinfo {title} {Introduction to the theory
  of disordered systems, wiley},}\ }\href@noop {} {\bibfield  {journal}
  {\bibinfo  {journal} {New York}\ } (\bibinfo {year} {1988})}\BibitemShut
  {NoStop}%
\bibitem [{\citenamefont {Landau}\ and\ \citenamefont
  {Lifshitz}(1960)}]{landau1960course}%
  \BibitemOpen
  \bibfield  {author} {\bibinfo {author} {\bibfnamefont {LD}~\bibnamefont
  {Landau}}\ and\ \bibinfo {author} {\bibfnamefont {EM}~\bibnamefont
  {Lifshitz}},\ }\href@noop {} {\emph {\bibinfo {title} {Course of theoretical
  physics. vol. 1: Mechanics}}}\ (\bibinfo  {publisher} {Oxford},\ \bibinfo
  {year} {1960})\BibitemShut {NoStop}%
\end{thebibliography}
\end{document}